\documentclass[aps,prd,draft,showpacs]{revtex4}

\begin{document}

\title{Renormalization Group Functions for the Radiative Symmetry Breaking Scheme
with Multi Mass Scale}
\author{Chungku Kim}
\date{\today}

\begin{abstract}
We obtain the renormalization group(RG) functions for the $O(N)$ scalar
field theory and the Higgs-Yukawa field theory with the Coleman-Weinberg
mechanism in which the symmetry breaking occurs radiatively by using the
method proposed previously in case of the neutral scalar field theory.
\end{abstract}
\maketitle


\affiliation{Department of Physics, Keimyung University, Daegu 704-701, KOREA%
}

\setcounter{page}{1} The electroweak(EW) model with the radiative symmetry
breaking(RSB) scheme \cite{CW} in which the scalar field does not have a
tree level mass and the symmetry breaking(SB) occurs from the effective
potential \cite{EP} has been studied extensively \cite{CWHiggs} due to its
predictive power for the magnitude of the Higgs boson mass. In order to
investigate the higher-order perturbative corrections, we need the
renormalization group(RG) function \cite{RGfn} in case of the RSB scheme.
Recently, we have obtained the RG function for the neutral scalar field
theory with the RSB scheme \cite{Kim} by two different procedures. Among
these, the one step procedure can be used in case of the theory which have
an effective potential depending on multi-mass scales \cite{multi}. In this
procedure, we first obtain the effective potential for the RSB scheme by a
finite transform of both the coupling constant and the classical field from
the well known effective potential in the minimal subtraction(MS) scheme.
Then we obtain the resulting change of the RG functions due to these
transforms.

In this paper, we will apply the one step procedure proposed in \cite{Kim}
to the case of the O(N) scalar field theory and the Higgs-Yukawa field
theory which have an effective potential depending on multi-mass scales and
have obtained the two-loop effective potential and the corresponding RG
functions in the RSB scheme.

\subsection{O(N) Scalar Field Theory}

The effective potential of the O(N) scalar field theory in the MS scheme has
the form \cite{Jones1} 
\begin{eqnarray}
V^{MS}(\lambda ,\phi ,\mu ) &=&\phi ^{4}[\frac{1}{24}\lambda +\kappa \lambda
^{2}(\frac{1}{16}L_{2}+\frac{n}{144}L_{1}-\frac{3}{32}-\frac{n}{96})+\kappa
^{2}\lambda ^{3}(\frac{3}{32}L_{2}^{2}+\frac{n}{48}L_{2}L_{1}+\frac{n^{2}}{%
864}L_{1}^{2}-(\frac{5}{16}+\frac{5n}{144})L_{2}  \nonumber \\
&&-(\frac{13n}{432}+\frac{n^{2}}{432})L_{1}+\frac{11}{32}+\frac{1}{2}\Omega
(1)+\frac{29n}{432}+\frac{5n}{54}\Omega (3)+\frac{n^{2}}{864})+..],
\end{eqnarray}
where $\kappa =(16\pi ^{2})^{-1}$, $n=N-1$ , $l$ is the order of the loop
and 
\begin{equation}
L_{1}\equiv \log \left( \frac{\lambda \phi ^{2}}{6\mu ^{2}}\right) \text{
and }L_{2}\equiv \log \left( \frac{\lambda \phi ^{2}}{2\mu ^{2}}\right)
\end{equation}
and $\Omega (x)$ is given by 
\begin{equation}
\Omega (x)=\frac{\sqrt{x(4-x)}}{x+2}\int_{0}^{\theta }\ln (2\sin t)dt\text{
(sin}\theta =\frac{\sqrt{x}}{2})
\end{equation}
Since the effective potential is independent of the renormalization mass
scale $\mu ,$ it satisfies the renormalization group equation

\begin{equation}
\lbrack \mu \frac{\partial }{\partial \mu }+\beta _{MS}(\lambda )\frac{%
\partial }{\partial \lambda }+\gamma _{MS}(\lambda )\phi \frac{\partial }{%
\partial \phi }]V^{MS}(\lambda ,\phi ,\mu )=0
\end{equation}
where the RG functions $\beta _{MS}$ and $\gamma _{MS}$ are given by\cite
{RGfn}

\begin{equation}
\beta ^{MS}(\lambda )=\mu \frac{d\lambda }{d\mu }=\kappa (3+\frac{n}{3}%
)\lambda ^{2}-\kappa ^{2}(\frac{17}{3}+n)\lambda ^{3}+\cdot \cdot \cdot
\equiv \sum_{l=1}^{\infty }b_{l}^{MS}\kappa ^{l}\lambda ^{l+1}
\end{equation}
and

\begin{equation}
\gamma ^{MS}(\lambda )=\frac{\mu }{\phi }\frac{d\phi }{d\mu }=-\kappa ^{2}(%
\frac{1}{12}+\frac{n}{36})\lambda ^{2}+\cdot \cdot \cdot \equiv
\sum_{l=1}^{\infty }g_{l}^{MS}\kappa ^{l}\lambda ^{l}
\end{equation}
In order to obtain the effective potential for the RSB scheme from $%
V^{MS}(\lambda ,\phi ,\mu )$, let us perform a finite transformations of the
coupling constant $\lambda $ and the classical field $\phi $ to $\widetilde{%
\lambda }$ and $\widetilde{\phi }$ as $\lambda =\lambda (\widetilde{\lambda }%
)$ and $\phi =\widetilde{\phi }$ $f(\widetilde{\lambda }$ $)$ respectively
so that the resulting effective potential for the RSB scheme $V^{R}(%
\widetilde{\lambda },\widetilde{\phi },\mu )$ has the form 
\begin{equation}
V^{R}(\widetilde{\lambda },\widetilde{\phi },\mu )=V^{MS}(\lambda (%
\widetilde{\lambda }),\widetilde{\phi }f(\widetilde{\lambda }),\mu
)=\sum_{l=0}^{\infty }\kappa ^{l}\widetilde{\phi }^{4}\sum_{n=0}^{l}a_{l,n}(%
\widetilde{\lambda })\log ^{n}\left( \frac{\widetilde{\phi }^{2}}{\mu ^{2}}%
\right) \equiv \sum_{l=0}^{\infty }\kappa ^{l}V_{l}^{R}(\widetilde{\lambda },%
\widetilde{\phi },\mu )
\end{equation}
and satisfies the renormalization condition 
\begin{equation}
\left[ \frac{d^{4}V^{R}(\widetilde{\lambda },\widetilde{\phi },\mu )}{d%
\widetilde{\phi }^{4}}\right] _{\widetilde{\phi }=\mu }=\widetilde{\lambda }.
\end{equation}
Then the corresponding change of the RG function due to these
transformations are given by \cite{Kim} 
\begin{equation}
\beta ^{R}(\widetilde{\lambda })=\mu \frac{\partial \widetilde{\lambda }}{%
\partial \mu }=\frac{\partial \widetilde{\lambda }}{\partial \lambda }\mu 
\frac{\partial \lambda }{\partial \mu }=\frac{\beta ^{MS}(\lambda (%
\widetilde{\lambda }))}{\frac{\partial \lambda (\widetilde{\lambda })}{%
\partial \widetilde{\lambda }}}\equiv \sum_{l=1}^{\infty }\kappa
^{l}b_{l}^{R}(\widetilde{\lambda }).
\end{equation}
and 
\begin{eqnarray}
\gamma ^{R}(\widetilde{\lambda }) &=&\frac{\mu }{\widetilde{\phi }}\frac{d%
\widetilde{\phi }}{d\mu }=\frac{1}{\widetilde{\phi }}[\frac{\mu }{f(%
\widetilde{\lambda })}\frac{\partial \phi }{\partial \mu }+\phi \mu \frac{%
\partial \widetilde{\lambda }}{\partial \mu }\frac{\partial f^{-1}(%
\widetilde{\lambda })}{\partial \widetilde{\lambda }}]  \nonumber \\
&=&\gamma ^{MS}(\lambda (\widetilde{\lambda }))+\beta ^{R}(\widetilde{%
\lambda })f(\widetilde{\lambda })\frac{\partial f^{-1}(\widetilde{\lambda })%
}{\partial \widetilde{\lambda }}\equiv \sum_{l=1}^{\infty }\kappa
^{l}g_{l}^{R}(\widetilde{\lambda }).
\end{eqnarray}
Note that the $l$-loop of order transformation $\lambda =\lambda (\widetilde{%
\lambda })$ and $\phi =\widetilde{\phi }$ $f(\widetilde{\lambda }$ $)$
determine the $(l+1)$-loop order coefficient of the RG functions $\beta _{R}(%
\widetilde{\lambda })$ and $\gamma _{R}(\widetilde{\lambda })$.

Now, let us apply the above procedure to obtain the two-loop effective
potential and the RG functions in the RSB scheme. The general form of the $%
O(\kappa )$ transforms can be written as 
\begin{eqnarray}
\lambda (\widetilde{\lambda }) &=&\widetilde{\lambda }+\kappa \widetilde{%
\lambda }^{2}(c_{110}l_{1}+c_{101}l_{2}+c_{100})  \nonumber \\
f(\widetilde{\lambda }) &=&1+\kappa \widetilde{\lambda }%
(d_{110}l_{1}+d_{101}l_{2}+d_{100})
\end{eqnarray}
where $l_{1}\equiv \log (\frac{\widetilde{\lambda }}{6})$ and $l_{2}\equiv
\log (\frac{\widetilde{\lambda }}{2})$ . By demanding that the one-loop
effective potential have the form (7), we obtain 
\begin{equation}
c_{110}=-4d_{110}-\frac{n}{6}\text{ , }c_{101}=-4d_{101}-\frac{3}{2}
\end{equation}
By substituting Eq.(11) into Eq.(10) we can see that if $d_{110},d_{101}$
and $d_{100}$ are not zero then the $g_{R}^{(2)}(\widetilde{\lambda })$
which is the$\ O(\kappa ^{2})$ coefficient of the RG function $\gamma ^{R}(%
\widetilde{\lambda })$ depend on $l_{1}$ and $l_{2}.$ Hence, in order to
obtain simple results for the two-loop RG functions, we choose $%
d_{110}=d_{101}=d_{100}=0$. Finally, by demanding that $V^{R}(\widetilde{%
\lambda },\widetilde{\phi },\mu )$ satisfies the renormalization condition
given in Eq.(8), we obtain 
\begin{equation}
\text{ }c_{100}=-4(1+\frac{n}{9})
\end{equation}
Next, let us write the $O(\kappa ^{2})$ transformation as 
\begin{eqnarray}
\lambda (\widetilde{\lambda }) &=&\widetilde{\lambda }-\kappa \widetilde{%
\lambda }^{2}(\frac{3}{2}l_{1}+\frac{n}{6}l_{2}+4(1+\frac{n}{9}))+\kappa ^{2}%
\widetilde{\lambda }%
^{3}(c_{220}l_{1}^{2}+c_{211}l_{1}l_{2}+c_{202}l_{2}^{2}+c_{210}l_{1}+c_{201}l_{2}+c_{200})+...
\nonumber \\
f(\widetilde{\lambda }) &=&1+\kappa ^{2}\widetilde{\lambda }%
^{2}(d_{220}l_{1}^{2}+d_{211}l_{1}l_{2}+d_{202}l_{2}^{2}+d_{210}l_{2}+d_{201}l_{2}+d_{200})+...
\end{eqnarray}
By substituting Eqs.(14) into (7) we can see that by choosing 
\begin{eqnarray}
c_{220} &=&\frac{n^{2}}{36}-4d_{220},\text{ }c_{211}=\frac{n}{2}-4d_{211},%
\text{ }c_{202}=\frac{9}{4}-4d_{202}\text{ },  \nonumber \\
c_{210} &=&-4d_{210}+\frac{4n^{2}}{27}+\frac{14n}{9}\ \text{and }%
c_{201}=-4d_{201}+\frac{5n}{3}+15
\end{eqnarray}
we can make those terms that depend on $l_{1}^{2},l_{1}l_{2},$ $l_{2}^{2},$ $%
l_{1}$ and $l_{2}$ in the two-loop effective potential vanish. Moreover, by
investigating $b_{R}^{(3)}(\widetilde{\lambda })$ and $g_{R}^{(3)}(%
\widetilde{\lambda })$ which is the three-loop coefficients of the RG
functions $\beta ^{R}$ and $\gamma ^{R}$, we can see that those terms that
depend on $l_{1}^{2},l_{1}l_{2}$ and $l_{2}^{2}$ can be made to zero by
choosing $d_{220}=d_{211}=d_{202}=0$ and that those terms that depend on $%
l_{1}$ and $l_{2}$ does not vanish by any choice of $d_{210}$ and $d_{201}$.
This means that the RG functions becomes a polynomial of the coupling
constant only up to the $O(\kappa ^{2}).$ Finally, by demanding the
renormalization condition given in Eq.(8), we obtain 
\begin{equation}
c_{200}=-4d_{200}-\frac{20n\Omega (2)}{9}+\frac{202n}{27}+\frac{139}{4}+%
\frac{113n^{2}}{324}-12\Omega (1)
\end{equation}
As a result, up to two loop order, we obtain the RG functions and effective
potential in RSB scheme as 
\begin{equation}
\beta ^{R}(\widetilde{\lambda })=\kappa (3+\frac{n}{3})\widetilde{\lambda }%
^{2}+\kappa ^{2}(-\frac{7}{6}+\frac{1}{18}n^{2})\widetilde{\lambda }%
^{3}+\cdot \cdot \cdot
\end{equation}
\begin{equation}
\gamma ^{R}(\widetilde{\lambda })=-\kappa ^{2}(\frac{1}{12}+\frac{n}{36})%
\widetilde{\lambda }^{2}+\cdot \cdot \cdot
\end{equation}
and 
\begin{eqnarray}
V^{R}(\widetilde{\lambda },\widetilde{\phi },\mu ) &=&\widetilde{\phi }^{4}[%
\frac{1}{24}\widetilde{\lambda }+\kappa \widetilde{\lambda }^{2}\{(\frac{1}{%
16}+\frac{n}{144})L-\frac{25}{96}-\frac{25n}{864}\}+\kappa ^{2}\widetilde{%
\lambda }^{3}\{(\frac{3}{32}+\frac{n}{48}+\frac{n^{2}}{864})L^{2}-(\frac{13}{%
16}+\frac{19n}{108}+\frac{11n^{2}}{1296})L  \nonumber \\
&&+\frac{55}{24}+\frac{635n}{1296}+\frac{85n^{2}}{3888}\}+O(\kappa ^{3})].
\end{eqnarray}
Note that up to this order $V^{R}(\widetilde{\lambda },\widetilde{\phi },\mu
)$ coincide with the effective potential given in Ref.[8] and satisfies the
RG equation 
\begin{equation}
\lbrack \mu \frac{\partial }{\partial \mu }+\beta ^{R}(\widetilde{\lambda })%
\frac{\partial }{\partial \widetilde{\lambda }}+\gamma ^{R}(\widetilde{%
\lambda })\widetilde{\phi }\frac{\partial }{\partial \widetilde{\phi }}%
]V^{R}(\widetilde{\lambda },\widetilde{\phi },\mu )=0
\end{equation}

\subsection{ Higgs-Yukawa field theory}

The effective potential of the Higgs-Yukawa field theory which is the O(4)
symmetric scalar field theory coupled to the Dirac fermion with the Yukawa
coupling constant $h$ in the MS scheme is given by \cite{Jones2} : 
\begin{eqnarray}
V^{MS}(\lambda ,h,\phi ,\mu ) &=&\phi ^{4}[\frac{1}{24}\lambda +\kappa (%
\frac{\lambda ^{2}}{48}\log (\frac{G}{\mu ^{2}})+\frac{\lambda ^{2}}{16}\log
(\frac{H}{\mu ^{2}})-\frac{3h^{2}}{4}\log (\frac{T}{\mu ^{2}})-\frac{\lambda
^{2}}{8}+\frac{9h^{2}}{8})+\kappa ^{2}[-\frac{\lambda ^{2}\phi ^{2}}{12}(%
\widehat{I}(H,H,H)  \nonumber \\
&&+\widehat{I}(G,H,H))+\frac{\lambda }{8}(\widehat{J}(H,H)+\widehat{J}(H,G)+%
\widehat{J}(G,G))+3h^{2}((2T-\frac{H}{2})\widehat{I}(H,T,T)  \nonumber \\
&&-\frac{T}{2}\widehat{I}(G,T,T)+(T-G)\widehat{I}(G,T,0)+\widehat{J}(T,T)-%
\widehat{J}(H,T)-2\widehat{J}(G,T)]+O(\kappa ^{3}),
\end{eqnarray}
where 
\begin{equation}
H=\frac{\lambda \phi ^{2}}{2},G=\frac{\lambda \phi ^{2}}{6},T=\frac{%
h^{2}\phi ^{2}}{2}
\end{equation}
and 
\begin{eqnarray}
\widehat{I}(x,y,z) &=&2(x\log (\frac{x}{\mu ^{2}})+y\log (\frac{y}{\mu ^{2}}%
)+z\log (\frac{z}{\mu ^{2}}))-\frac{1}{2}(y+z-x)\log (\frac{y}{\mu ^{2}}%
)\log (\frac{z}{\mu ^{2}})-\frac{1}{2}(z+x-y)\log (\frac{z}{\mu ^{2}})\log (%
\frac{x}{\mu ^{2}})  \nonumber \\
&&-\frac{1}{2}(x+y-z)\log (\frac{x}{\mu ^{2}})\log (\frac{y}{\mu ^{2}})-%
\frac{5}{2}(x+y+z)-\frac{1}{2}\xi (x,y,z)
\end{eqnarray}
and 
\begin{equation}
\widehat{J}(x,y)=xy(1-\log (\frac{x}{\mu ^{2}})-y\log (\frac{y}{\mu ^{2}}%
)+\log (\frac{x}{\mu ^{2}})\log (\frac{y}{\mu ^{2}}))
\end{equation}
The $\xi (x,y,z)$ is a function independent of the logarithms and the
detailed form of this function is given in \cite{Jones2}. Since the
effective potential is independent of the renormalization mass scale $\mu ,$
it satisfies the renormalization group equation

\begin{equation}
\lbrack \mu \frac{\partial }{\partial \mu }+\beta _{\lambda }^{MS}\frac{%
\partial }{\partial \lambda }+\beta _{h}^{MS}\frac{\partial }{\partial h}%
+\gamma ^{MS}(\lambda )\phi \frac{\partial }{\partial \phi }]V^{MS}(\lambda
,h,\phi ,\mu )=0
\end{equation}
where the RG functions $\beta _{MS}$ and $\gamma _{MS}$ are given by\cite
{RGfn}

\begin{eqnarray}
\beta _{\lambda }^{MS} &=&\mu \frac{d\lambda }{d\mu }=\kappa (4\lambda
^{2}+12\lambda h^{2}-36h^{4})+\kappa ^{2}(-\frac{26}{3}\lambda
^{3}-24\lambda ^{2}h^{2}-3\lambda h^{4}+180h^{6})+... \\
\beta _{h^{2}}^{MS} &=&\mu \frac{dh^{2}}{d\mu }=9\kappa h^{4}+\kappa ^{2}(%
\frac{1}{3}\lambda ^{2}h^{2}-4\lambda h^{4}-24h^{6})+...
\end{eqnarray}
and

\begin{equation}
\gamma ^{MS}=\frac{\mu }{\phi }\frac{d\phi }{d\mu }=-3\kappa h^{2}+\kappa
^{2}(-\frac{1}{6}\lambda ^{2}+\frac{27}{4}h^{4})\cdot \cdot \cdot
\end{equation}
Since the effective potential depends on two different coupling constants $%
\lambda $ and $h$, let us generalize the previous results (Eqs.(9) and (10))
which correspond to a single coupling constant $\lambda $ to the case where
several coupling constants $\lambda _{i}$ ($i=1,..,N)$ exist and consider
the finite transforms of the coupling constants $\lambda _{i}$ and the
classical field $\phi $ as 
\begin{equation}
\lambda _{i}=\lambda _{i}(\widetilde{\lambda }_{j})\text{ }(i,j=1,..,p)
\end{equation}
and 
\begin{equation}
\phi =\widetilde{\phi }\text{ }f(\widetilde{\lambda }_{j})\text{ }
\end{equation}
Then the RG of the coupling constants in RSB scheme can be obtained as 
\begin{equation}
\beta _{i}^{R}(\widetilde{\lambda }\text{ })=\mu \frac{d\widetilde{\lambda }%
_{i}}{d\mu }=\sum_{j=1}^{p}\frac{\partial \widetilde{\lambda }_{i}}{\partial
\lambda _{j}}\mu \frac{d\lambda _{j}}{d\mu }=
\sum_{j=1}^{p}(M^{-1})_{ij}\beta _{j}^{MS}(\lambda (\widetilde{\lambda }%
\text{ }))\equiv\sum_{l=1}^{\infty }\kappa ^{l}b_{l}^{R}(\widetilde{\lambda }%
).
\end{equation}
where $(M)_{ij}\equiv \frac{\partial \lambda _{i}}{\partial \widetilde{%
\lambda }_{i}}$ which can be obtained from (29). The RG of the classical
field $\phi $ in RSB\ scheme can be obtained by a simple extension of the
Eq.(10) as 
\begin{equation}
\gamma ^{R}(\widetilde{\lambda })=\mu \frac{d\widetilde{\phi }}{d\mu }%
=\gamma ^{MS}(\lambda (\widetilde{\lambda }))+\sum_{i=1}^{p}\beta _{i}^{R}(%
\widetilde{\lambda }\text{ })f(\widetilde{\lambda })\frac{\partial f^{-1}(%
\widetilde{\lambda })}{\partial \widetilde{\lambda }_{i}}\equiv
\sum_{l=1}^{\infty }\kappa ^{l}g_{l}^{R}(\widetilde{\lambda }).
\end{equation}
As in case of the O(N) symmetric scalar field theory, we have searched the
finite transformation of the coupling constants and the classical field by
requiring that the one-loop effective potential take the form eq.(7),
satisfy the renormalization condition given in eq.(8) and that the
coefficients of the logarithms $L\equiv \log \left( \frac{\widetilde{\phi }%
^{2}}{\mu ^{2}}\right) $of the two loop effective potential does not depend
on $\log \left( \widetilde{\lambda }\right) $ and $\log (\widetilde{h}^{2}).$
The resulting one-loop order transformation was not unique and one possible
solution was 
\begin{eqnarray}
\lambda &=&\widetilde{\lambda }+\kappa [(-\frac{3}{2}l_{1}-\frac{1}{2}l_{2}-%
\frac{16}{3})\widetilde{\lambda }^{2}+(18l_{3}+48)\widetilde{h}^{4}] 
\nonumber \\
h^{2} &=&\widetilde{h}^{2}+\kappa [(\frac{1}{4}l_{1}+\frac{1}{12}l_{2}-\frac{%
1}{6})\widetilde{\lambda }^{2}+(-\frac{3}{2}l_{1}-\frac{1}{2}l_{2}+2l_{3})
\widetilde{\lambda }\widetilde{h}^{2}-\frac{3}{2}\widetilde{h}^{4}]  \nonumber
\\
f &=&1
\end{eqnarray}
where $l_{1}\equiv \log (\frac{\widetilde{\lambda }}{6})$ , $l_{2}\equiv
\log (\frac{\widetilde{\lambda }}{2})$ and $l_{3}\equiv \log (\frac{%
\widetilde{h}^{2}}{2})$. Then, by using the equations (21) and (30)-(33),
the effective potential and the RG functions in the RSB scheme can be
determined up to the two loop order as 
\begin{eqnarray}
V^{R}(\widetilde{\lambda },\widetilde{h},\widetilde{\phi },\mu ) &=&\phi
^{4}[\frac{1}{24}\widetilde{\lambda }+\kappa (\frac{\widetilde{\lambda }^{2}%
}{12}L-\frac{3}{4}\widetilde{h}^{2}L-\frac{25}{72}\widetilde{\lambda }^{2}+%
\frac{25}{8}\widetilde{h}^{2})+\kappa ^{2}[(\frac{1}{6}\widetilde{\lambda }%
^{3}+\frac{1}{4}\widetilde{\lambda }^{2}\widetilde{h}^{2}-\frac{3}{2}%
\widetilde{\lambda }\widetilde{h}^{4}-\frac{9}{8}\widetilde{h}^{6})L^{2} 
\nonumber \\
&&+(-\frac{17}{12}\widetilde{\lambda }^{3}-\frac{3}{4}\widetilde{\lambda }%
^{2}\widetilde{h}^{2}+\frac{23}{2}\widetilde{\lambda }\widetilde{h}^{4}+6%
\widetilde{h}^{6})L+\frac{25}{24}\widetilde{\lambda }^{3}+\frac{5}{24}%
\widetilde{\lambda }^{2}\widetilde{h}^{2}-\frac{365}{12}\widetilde{\lambda }%
\widetilde{h}^{4}-\frac{95}{8}\widetilde{h}^{6}]+O(\kappa ^{3}),
\end{eqnarray}
\begin{eqnarray}
\beta _{\widetilde{\lambda }}^{R} &=&\mu \frac{\partial \widetilde{\lambda }%
}{\partial \mu }=\kappa (4\widetilde{\lambda }^{2}+12\widetilde{\lambda }%
\widetilde{h}^{2}-36\widetilde{h}^{4})+\kappa ^{2}[-\frac{8}{3}\widetilde{%
\lambda }^{3}+76\widetilde{\lambda }^{2}\widetilde{h}^{2}-75\widetilde{%
\lambda }\widetilde{h}^{4}-2706\widetilde{h}^{6}+(3\widetilde{\lambda }%
^{3}-18\widetilde{\lambda }^{2}\widetilde{h}^{2})l_{1}  \nonumber \\
&&+(\widetilde{\lambda }^{3}-6\widetilde{\lambda }^{2}\widetilde{h}%
^{2})l_{2}+(24\widetilde{\lambda }^{2}\widetilde{h}^{2}-18\widetilde{\lambda 
}\widetilde{h}^{4})l_{3}],
\end{eqnarray}
\begin{eqnarray}
\beta _{\widetilde{h}^{2}}^{R} &=&\mu \frac{\partial \widetilde{h}^{2}}{%
\partial \mu }=9\kappa \widetilde{h}^{4}+\kappa ^{2}[\frac{16}{3}\widetilde{%
\lambda }^{2}\widetilde{h}^{2}+2\widetilde{\lambda }\widetilde{h}^{4}-\frac{%
165}{2}\widetilde{h}^{6}+(-2\widetilde{\lambda }^{3}+\frac{9}{2}\widetilde{%
\lambda }^{2}\widetilde{h}^{2}+\frac{45}{2}\widetilde{\lambda }\widetilde{h}%
^{4}-54\widetilde{h}^{6})l_{1}  \nonumber \\
&&+(-\frac{2}{3}\widetilde{\lambda }^{3}+\frac{3}{2}\widetilde{\lambda }^{2}%
\widetilde{h}^{2}+\frac{15}{2}\widetilde{\lambda }\widetilde{h}^{4}-18%
\widetilde{h}^{6})l_{2}+(-8\widetilde{\lambda }^{2}\widetilde{h}^{2}-6%
\widetilde{\lambda }\widetilde{h}^{4}+72\widetilde{h}^{6})l_{3}]
\end{eqnarray}
and 
\begin{equation}
\gamma ^{R}(\widetilde{\lambda },\widetilde{h})=\mu \frac{\partial 
\widetilde{\lambda }}{\partial \mu }=-3\kappa \widetilde{h}^{2}+\kappa ^{2}[%
\frac{1}{3}\widetilde{\lambda }^{2}+\frac{27}{4}\widetilde{h}^{4}+(-\frac{3}{%
4}\widetilde{\lambda }^{2}+\frac{9}{2}\widetilde{\lambda }\widetilde{h}%
^{2})l_{1}+(-\frac{1}{4}\widetilde{\lambda }^{2}+\frac{3}{2}\widetilde{%
\lambda }\widetilde{h}^{2})l_{2}+(-6\widetilde{\lambda }\widetilde{h}^{2}+%
\frac{9}{2}\widetilde{h}^{4})l_{3}].
\end{equation}
where the $L$ independent term of the $O(\kappa ^{2})$ term of $V^{R}(%
\widetilde{\lambda },\widetilde{h},\widetilde{\phi },\mu )$ was fixed by the
renormalization condition given in (8). One can check that up to $O(\kappa
^{2}),$ $V^{R}(\widetilde{\lambda },\widetilde{h},\widetilde{\phi },\mu )$
satisfies the RG equation 
\begin{equation}
\lbrack \mu \frac{\partial }{\partial \mu }+\beta ^{R}(\widetilde{\lambda })%
\frac{\partial }{\partial \widetilde{\lambda }}+\beta ^{R}(\widetilde{h})%
\frac{\partial }{\partial \widetilde{h}}+\gamma ^{R}(\widetilde{\lambda })%
\widetilde{\phi }\frac{\partial }{\partial \widetilde{\phi }}]V^{R}(%
\widetilde{\lambda },\widetilde{h},\widetilde{\phi },\mu )=0
\end{equation}
Form this equation, one can see that the coefficient of the $L$ in the $%
O(\kappa ^{2})$ term of $V^{R}(\widetilde{\lambda },\widetilde{h},\widetilde{%
\phi },\mu )$ is related to the combination $b_{2}^{R}(\widetilde{\lambda }%
)+4g_{2}^{R}(\widetilde{\lambda }).$ Hence, although both $b_{2}^{R}(%
\widetilde{\lambda })$ and $g_{2}^{R}(\widetilde{\lambda })$ depend on $l_{i}
$ terms, the combination $b_{2}^{R}(\widetilde{\lambda })+4g_{2}^{R}(%
\widetilde{\lambda })$ does not depend on $l_{i}$ terms so that the
coefficient of the $L$ in the $O(\kappa ^{2})$ term of $V^{R}(\widetilde{%
\lambda },\widetilde{h},\widetilde{\phi },\mu )$ is independent of $l_{i}$
terms. In the case of the Higgs-Yukawa theory which contains three different
mass scales in the effective potential, one can check that the $l_{i}$ terms
always appear in the $O(\kappa ^{2})$ RG functions even with the most
general form of the finite transformation of the coupling constants and the
classical field. The appearance of the logarithms in the coefficients of the RG
functions also happens in case of the multi-scale renormalization\cite{wiesen}.
As a result of these logarithms, the RG running of the coupling constants and the classical
field will be different compared to the case of the usual RG functions which
is a polynomial of the coupling constants and this is in progress. 
\center{ACKNOWLEDGMENTS} \endcenter

This research was supported by the Institute of the Basic Science Research
Center.

\end{document}